\documentclass{article}

\usepackage{microtype}
\usepackage{graphicx}
\usepackage{subfigure}
\usepackage{booktabs} 
\usepackage{array}

\usepackage{hyperref}
\usepackage{enumitem}

\usepackage[accepted]{icml2024}

\usepackage{amsmath}
\usepackage{amssymb}
\usepackage{mathtools}
\usepackage{amsthm}

\usepackage[capitalize,noabbrev]{cleveref}

\theoremstyle{plain}
\newtheorem{theorem}{Theorem}[section]

\theoremstyle{definition}

\theoremstyle{remark}

\usepackage[textsize=tiny]{todonotes}

\icmltitlerunning{Policy Gradients for Optimal Parallel Tempering MCMC}

\begin{document}

\twocolumn[
\icmltitle{Policy Gradients for Optimal Parallel Tempering MCMC}



\icmlsetsymbol{equal}{*}

\begin{icmlauthorlist}
\icmlauthor{Daniel Zhao}{yyy}
\icmlauthor{Natesh S. Pillai}{yyy}
\end{icmlauthorlist}

\icmlaffiliation{yyy}{Department of Statistics, Harvard University, Cambridge, USA}

\icmlcorrespondingauthor{Daniel Zhao}{danielzhao@alumni.harvard.edu}

\icmlkeywords{reinforcement learning, MCMC, policy gradient, Parallel tempering, ICML}

\vskip 0.3in
]



\printAffiliationsAndNotice{}

\begin{abstract}
Parallel tempering is a meta-algorithm for Markov Chain Monte Carlo that uses multiple chains to sample from tempered versions of the target distribution, enhancing mixing in multi-modal distributions that are challenging for traditional methods. The effectiveness of parallel tempering is heavily influenced by the selection of chain temperatures. Here, we present an adaptive temperature selection algorithm that dynamically adjusts temperatures during sampling using a policy gradient approach. Experiments demonstrate that our method can achieve lower integrated autocorrelation times compared to traditional geometrically spaced temperatures and uniform acceptance rate schemes on benchmark distributions.
\end{abstract}

\section{Introduction}

The core of many Bayesian inference problems is the challenge of estimating model parameters when the underlying probability distribution is too complex for direct solution or sampling. Markov Chain Monte Carlo (MCMC) methods address this problem by generating correlated random samples to approximate the distribution. However, traditional MCMC algorithms frequently face difficulties when sampling from target distributions that are highly multimodal or possess rugged energy landscapes. In such cases, chains can become trapped in local modes, leading to poor exploration of the space and slow convergence rates.

To mitigate this problem, parallel tempering MCMC introduces auxiliary tempered chains which exchange information by periodically swapping states with their neighbors. While low temperature chains are prone to becoming trapped into local energy minima, hot chains can more easily traverse entropic barriers in the flattened energy landscape to jump between modes. Having adjacent chains swap states allows the colder chain to explore new modes which would otherwise take much longer to reach. However, determining the optimal temperatures to maximize the efficiency of parallel tempering remains an open problem since the temperatures at which a chain can effectively cross entropic barriers (e.g. near a phase transition) varies widely according to the target distribution. Adaptive approaches which can automatically adjust the temperature ladder to the features of the target distribution have thus become popular in recent works \cite{Katzgraber_2006, miasojedow2012adaptive, Vousden_2015}. 

In this paper, we introduce a novel reinforcement learning approach to optimizing parallel tempering Markov Chain Monte Carlo (MCMC). We conceptualize the selection of the temperature ladder as a stateless policy optimization problem with an associated reward function. The term "stateless" reflects that updating the temperature ladder does not alter the chain's positions. The reward function is designed to measure the efficiency of the sampler, such that optimizing this reward enhances the mixing of the Markov chain by minimizing sample autocorrelation. Although most adaptive algorithms in the literature assume that uniform acceptance rates between chains optimize sampler efficiency \cite{miasojedow2012adaptive, Vousden_2015}, we explore several alternative reward formulations in our approach.

\textbf{Our contributions} in this paper are:

\textbf{Novel algorithm for optimizing parallel tempering MCMC.} We introduce a policy gradient-based approach to the temperature ladder selection problem. Our algorithm gradually shapes the temperatures to maximize the long-term average reward. We show that, with diminishing update magnitudes, it satisfies the necessary ergodic properties to ensure convergence to the target distribution. Furthermore, our algorithm does not require separated train/sampling phases, due to its statelessness.

\textbf{Distance metric for estimating the efficiency of swaps.} We propose the use of a swap mean distance metric as the primary component of the reward function. Intuitively, for each swap attempt with the coldest chain, this metric measures how surprising the new swapped state is compared to its most recent states, measured by its mean distance. We show that the swap mean-distance metric is strongly correlated with ACT.

\textbf{Experimental findings.} We run experiments to demonstrate the effectiveness of our method compared to benchmarks in the literature. Our results show that the algorithm is stable and roughly replicates state-of-the-art results \cite{Vousden_2015}. Additionally, when optimizing for the swap mean-distance metric, we are able to achieve lower ACT than both geometrically spaced temperatures and uniform acceptance rate chains on test distributions. This suggests that incorporating both metrics in the objective function is better than achieving uniform acceptance rates alone.

\section{Related Work}\label{rw}

Many authors have advocated choosing a temperature ladder which achieves a fixed, uniform acceptance rate across all chains \cite{roberts1997, Sugita1999REMD, kofke2002, kone2005}. To achieve this, the prevailing heuristic set forth by Kofke et al. (2002) is geometrically spacing the temperatures so that the ratio of the temperatures of any two adjacent chains is constant. It has been shown that geometric spacing is optimal for the Gaussian distribution but not sufficient to achieve uniform acceptance rates in general \cite{Katzgraber_2006, Vousden_2015}.

\textbf{Adaptive Affine-Invariant Sampler}

Vousden et al. (2015) proposed an adaptive algorithm which aims to achieve uniform acceptance rates across chains \cite{Vousden_2015}. Their algorithm periodically updates the temperatures based on adjacent acceptance rates so that neighbors with low acceptance rates are moved closer together in temperature, while high acceptance rate neighbors are pushed apart. The maximum temperature is fixed to infinity so that the target distribution becomes flat in the hottest chain, bypassing the need to manually select a maximum temperature which could be insufficient to overcome pathological features in the distribution. Consequently, every update only modulates the temperatures of intermediate chains. The eventual acceptance rate, then, is dependent only on the chosen number of chains and the target distribution. The authors show that their method converges and empirically achieves lower ACT than the geometric temperature ladder in non-Gaussian cases. In this work, we use their implementation of parallel tempering MCMC, which uses as its base the ensemble affine invariant sampler \cite{goodman2010}. See Appendix A for details.

\textbf{Conceptual RL for MCMC}

Optimization problems in MCMC often display properties that make a reinforcement learning approach appealing. For example, Markovian sequential decision-making mirrors the typical settings in which RL excels, and agent-based learning methods are usually well suited to deal with the curse of dimensionality and the exploration-exploitation trade-off inherent in optimization problems. Some authors have begun to create conceptual frameworks for using reinforcement learning to improve MCMC sampling \cite{Bojesen_2018, Chung2020MultiagentRL, wang2024reinforcement}. For Metropolis-Hastings, the state of the MDP is identified with the state of the Markov chain, while the action is identified with the proposal mechanism. This naturally leads to a policy learning scenario where the objective is to learn a stochastic policy that represents the optimal proposal distribution for maximizing sampler efficiency. Empirically, a DDPG implementation of a policy gradient adaptive Metropolis-Hastings algorithm was shown to outperform existing benchmarks \cite{wang2024reinforcement}, demonstrating the potential of RL for optimizing MCMC.

\section{Problem Setup}

\subsection{Background}
Let $\mathcal{X}$ be the phase space and $f$ the target distribution. In parallel tempering MCMC, $M$ replica chains $\{(X^{(1)}_t)_{t \in \mathbb{N}},\ldots (X^{(M)}_t)_{t \in \mathbb{N}}\}$ run simultaneously at different temperatures $T = \{T_1,...,T_N\}$ where $T_1 < T_2 < ... < T_M$ and $T_1 = 1$. The inverse of the temperatures are the betas: $\beta_i = 1/{T_i}$, and the stationary distribution $f^{\beta_i}$ of $(X^{(i)})$ is defined as $$f^{\beta_i}(x^{(i)}) := (f(x^{(i)}))^{\beta_i} \hspace{8pt} \forall x^{(i)} \in \mathcal{X}.$$ The joint chain $(X^{(1)}_t \ldots X^{(M)}_t)_{t \in \mathbb{N}}$ is also a Markov chain, whose canonical distribution is $$\tilde{f}(x^{(1)},...,x^{(M)}) = f^{\beta_1}(x^{(1)}) \ldots f^{\beta_M}(x^{(M)}),$$ from which the coldest chain marginally has stationary distribution $f$. While the chains run independently, they will periodically attempt a Metropolis swap move where two adjacent chains $X^{(i)}_t$ and $X^{(i+1)}_t$ propose to switch states $(x, y) \mapsto (y,x)$. This move is accepted with probability
\begin{equation}\label{eq:swap}
\begin{split}
    A(x, y) = \min\left(1, \left(\frac{\pi(y)}{\pi(x)}\right)^{\beta_{i}} \left(\frac{\pi(x)}{\pi(y)}\right)^{\beta_{i+1}}\right)
\end{split}
\end{equation}
to preserve the detailed balance of the joint chain, where $\pi$ is the unnormalized density. 

Denote the acceptance rate $A_i$ of a chain as the proportion of attempted swaps between itself and any adjacent chains which are successful. 

\subsection{Maximizing Average Reward}

Suppose that each temperature ladder $T$ is associated with a reward distribution $R_T$ which represents the average reward obtained after running all chains some fixed $N$ steps, assuming stationarity. We consider the problem of selecting $$T^* = \text{argmax}_T \hspace{3pt} E[R_T].$$ In this setup, the action is identified with the temperature ladder $T$, and the state is null. The lack of a state presents a unique challenge compared to previous works in RL for adaptive Metropolis-Hastings.

This problem is also known as the infinite-horizon Lipschitz bandits problem, for which the finite-time case has been studied \cite{Agrawal-1995, magureanu2014lipschitz}. However, our goal is not to maximize the cumulative reward in a fixed number of samples, but to achieve convergence towards the optimal temperature configuration with the highest mean reward over time. This motivates gradient-based methods, which are guaranteed to converge to local minima. They are especially powerful since certain reward functions may further be convex. For example, the variance of all acceptance rates is convex, since the acceptance rate between two chains increases monotonically with the difference in their temperatures.

\subsection{Policy Gradient Methods}
Policy gradient methods are popular for problems where the action space is continuous and high-dimensional—such as temperature selection—because parameterizing the policy allows us to output a continuous action directly, rather than relying on action-value estimation as in value-based methods. In addition, far fewer parameters are needed to be learned for the policy, compared to the value function, which is often complex to approximate.

Given an MDP $(S, A, P_a, R_a)$ with states $S$, actions $A$, transition kernel $P_a$, and reward function $R_a$ associated with action $a$, policy gradient methods attempt to learn a parameterized policy mapping $\pi_\theta: S \times A \rightarrow [0,1]$ which maximizes the cumulative reward $G_L$ where $t = 1,\cdots,L$.

In the infinite time horizon context, we want $\pi_\theta$ which maximizes the expected reward at stationarity: define the objective function as $J(\theta) := E_{\pi_\theta}[R_a]$. Notice that in the context of parallel tempering MCMC, the Markov chain we refer to in this section corresponds to the sequence of temperatures $(T_i)_{i \leq L}$, and not the Markov chain of the sampler.

\section{Policy Gradients for Temperature Selection}

The temperature selection problem can be modeled as a single-state RL problem in which the temperatures $a_t$ are sampled from the stochastic policy $\pi_\theta$, parameterized by some $\theta$. If the density $\pi_\theta$ is centered around $\theta$, we have a natural interpretation of $\theta$ as the best estimate of the optimal action.

The action space is taken to be the space of log temperature differences. That is, for $D_i = \log \beta_i - \log \beta_{i+1}$, the space of configurations $A = (D_1,\ldots,D_{M-1})$ is $[0, \infty)^{M-1}$, which can further be made compact (and thus the reward function is Lipschitz continuous) by choosing reasonable bounds $A = [D_{\text{min}}, D_{\text{max}}]^{M-1}$ for constants $0 < D_{\text{min}} < D_{\text{max}}$. Considering the log-differences between temperatures ensures that the scale of all parameters are similar, which contributes to learning stability, and allows us to easily fix $\beta_1 = 1$ and $\beta_M = 0$. In our experiments, we take $D_{\text{min}} = 0.01$, $D_{\text{max}} = 10$, and $N = 500$.

Note that in the single-state case, the gradient of the objective function $J(\theta)$ reduces to
\begin{align*}
    \nabla_\theta J(\theta) &= \int_{A} \nabla_\theta \pi_\theta(a) E_\pi[R_t | A_t = a] \hspace{2pt} da \\
    &= \int_{A} \pi_\theta(a) \nabla_\theta \log \pi_\theta(a) E_\pi[R_t | A_t = a] \hspace{2pt} da.
\end{align*}
via the Policy Gradient Theorem (See Appendix C). Thus, we may obtain an unbiased Monte Carlo estimate of $\nabla_\theta J(\theta)$ by calculating $\nabla_\theta \log \pi_\theta(a) \hspace{2pt} E_\pi[R_t | A_t = a]$ for every action of the current policy $\pi_\theta$. The update is performed as the last step:
$$\theta_{t+1} = \theta_t + \alpha \nabla_\theta J(\theta) |_{\theta=\theta_n}$$ for learning rate $\alpha$.

In Algorithm \ref{sspg}, we describe our method generally, so that it remains applicable to other optimization problems in MCMC which may have similar setups.
\begin{algorithm}[H]
\caption{Single-State Policy Gradient for Maximizing Average Reward}
\begin{algorithmic}[1]
\INPUT Initialize policy parameters $\theta_0$, initial positions of walkers $\{\mathbf{x}_0^{(i)}\}_{i=1}^{M}$, policy distribution $\pi_\theta(\cdot)$
\FOR {$t = 1$ to $L$}
    \STATE Generate a set of sampler parameters $a_t$ by drawing from policy distribution $\pi_\theta$ \label{sspg}
    \STATE Run sampler with parameters $a_t$, observe reward $r_t$
    \STATE Normalize $r_t$ and use it to estimate the average reward $\bar{R}$ based on observed rewards
    \STATE Calculate the gradient of the policy $\nabla_{\theta} \log \pi_\theta(a_t)$ w.r.t. $\theta$ and clip it
    \STATE Update the policy parameters:\\ $\theta \leftarrow \theta + \alpha (r_t - \bar{R}) \nabla_{\theta} \log \pi_\theta(a_t)$
\ENDFOR
\OUTPUT Samples $\{\mathbf{x}_j^{(1)}\}_{j=1}^{L}$, Policy parameters $\theta_L$
\end{algorithmic}
\end{algorithm}
In our experiments, we take the policy function to be normal centered at $\theta$: $\pi_\theta(\cdot) \sim \mathcal{N}(\theta, \sigma I)$ so that the gradient is simply $$\nabla_{\theta} \log \pi_\theta(a_t) = -\sigma^{-1}(a_t - \theta).$$ The average reward $\bar{R}$ is estimated as the average of the last $500$ rewards. Initial $\theta_0$ values are spaced such that $\Delta \theta_i = 1$.

\subsection{Convergence Analysis}
In general, if the temperature ladder is changed during sampling, there is no guarantee that the coldest chain converges to the target distribution. However, it has been shown \cite{Roberts2007, saksman2010} that the convergence of an adaptive MCMC method is preserved if the following conditions hold.
\begin{theorem}
    Let $\{X_t\}_{t \in \mathbb{N}}$ be Markov chain and $P_{\Gamma_{t}}(x, \cdot)$ the $t^{\text{th}}$ adapted conditional density for $X_{t+1}$ given $X_t = x$. Then the Strong Law of Large Numbers holds for $\{X_t\}$ if two conditions are met:
    \begin{enumerate}[label=(\alph*)]
        \item (Diminishing Adaptation) For every starting $x \in \mathcal{X}$, $$\lim_{t\rightarrow \infty} \sup_{x\in \mathcal{X}} \| P_{\Gamma_{t+1}} (x, \cdot) - P_{\Gamma_{t}} (x, \cdot)\| = 0.$$
        \item (Containment) For any $\epsilon > 0$, the sequence $$\{M_\epsilon (X_t, \Gamma_t)\}_{t \in \mathbb{N}}$$ is bounded in probability, where $$M_\epsilon(x,\Gamma_t) := \inf \{n \geq 1: \| P^n_{\Gamma_t}(x, \cdot) - \pi(\cdot) \| \leq \epsilon\}.$$
    \end{enumerate}
\end{theorem}
The Diminishing Adaptation condition requires that changes to the transition kernel decay to zero, and Containment requires the convergence time of the chain to be bounded in probability—note that this is trivially satisfied with a compact phase space. Subsequent works extended Theorem 4.1 for weaker forms of the Containment condition \cite{bai2011, rosenthal2018}. 

\textbf{Diminishing Adaptation.}
To modify our algorithm to satisfy the Diminishing Adaptation condition, we artificially dampen the variance of the sampling from the policy $\pi_{\theta_t}(\cdot) \sim \mathcal{N}(\theta_t, \epsilon_t \sigma I)$ by decaying $\epsilon_t \rightarrow 0$. The gradient $\nabla_{\theta} \log \pi_\theta(T) = -\sigma^{-1}(T - \theta)$ is not scaled with $\epsilon_t$, so it also diminishes to $0$ as $t \rightarrow \infty$. 

This ensures that adaptations vanish over sufficiently long time scales, and all temperatures approach a fixed value. In fact, exponential decay of $\epsilon_t$ also ensures that the policy $\pi_{\theta_t}$ converges fairly quickly to $\theta_t$, so the optimal temperature ladder $T^*$ is solved for by estimating $\theta^*$. Not scaling the gradient does bias the estimates of $\nabla_\theta J(\theta)$, but our experiments show that it does not significantly affect the convergence of the algorithm.

\textbf{Containment.}
Our focus is on the convergence of the coldest chain to the target distribution. A sufficient condition for ensuring Containment is that the convergence time of the coldest chain remains bounded in probability. Since the temperature of this chain is fixed, failure to converge occurs only if the coldest Markov chain itself fails to converge. For most practical applications, this condition is quite weak and easily satisfied. Importantly, ensuring the ergodicity of the coldest chain is generally a matter of designing an effective underlying sampling algorithm, rather than a property dependent on the adaptive scheme.

\section{Swap Mean Distance}

With guarantees on our algorithm's convergence, the next critical step is selecting a reward function that effectively quantifies sampler efficiency. Efficiency is often assessed through the autocorrelation of successive samples, as highly autocorrelated Markov chains tend to mix slowly. In the context of parallel tempering, frequent state exchanges with hotter chains can mitigate autocorrelation, accelerating mixing. However, designing a feedback mechanism for the swap process that encapsulates this objective remains an unresolved challenge. In this section, we evaluate several candidate reward functions and examine their implications for mixing behavior. Additionally, we introduce a novel metric that leverages the distance between states involved in a swap to guide the optimization of mixing dynamics.

\textbf{Integrated Autocorrelation Time}

The most direct metric for measuring sampler efficiency is the integrated autocorrelation time (ACT). For any function $h$ suppose $\langle h^N \rangle$ is a Monte Carlo estimator for $E[h(X)]$. Its variance when the samples are correlated converges: $$Var[\langle h^N \rangle] \longrightarrow \dfrac{\tau_h}{N} \text{ } Var[h^\pi]$$ where $\tau_h$ is the integrated autocorrelation time of the Markov chain $(X_t)_{t \in \mathbb{N}}$, given by
\begin{equation}
    \tau_h = \sum_{i=-\infty}^\infty \rho_i.
\end{equation}
and $\rho_i = Cov[h(X_1), h(X_{1+i})]$ is the autocorrelation of the $i^{\text{th}}$ lag.

As $N$ is quite large in most MCMC applications, the asymptotic variance is usually approximated quite well by the ACT. Optimizing the efficiency of the sampler is then achieved by minimizing the integrated autocorrelation time.

In practice, estimating the autocorrelation time of a sampler is inconsistent with limited samples, so it is not necessarily a good feedback mechanism to use for distributions whose densities are not known a-priori.

\textbf{Uniform Acceptance Rate}

Achieving uniform acceptance rates across chains is a rather intuitive objective, as consistent swaps between chains allows information to be exchanged effectively down the entire temperature ladder. Empirical implementations has been shown to be more efficient than geometrically spaced temperatures \cite{Vousden_2015, Sugita1999REMD}. Various reward functions can be employed to optimize for uniform acceptance rates. For example, in our experiments, we find that maximizing the negative standard deviation of the acceptance rates achieves this goal.

\textbf{Expected Squared Jumping Distance}

Another metric for assessing sampler efficiency proposed by \cite{atchade2011} is the expected squared jumping distance (ESJD). Suppose we have some fixed chain at temperature $\beta$, a swap is proposed with some other chain at temperature $\beta + \epsilon.$ Let $\gamma = \beta + \epsilon$ if the swap is accepted, or $\gamma = \beta$ if the swap is rejected. Then, define
\begin{align*}
    ESJD_\beta &:= E_\pi[(\gamma - \beta)^2]\\
    &= \epsilon^2 E_\pi[A].
\end{align*}
The acceptance rate, as a swap-dependent metric, primarily quantifies the efficiency of the swap mechanism but does not capture the broader contribution of individual swaps to enhancing the overall mixing of the system. In contrast, information-theoretic measures, such as the KL-divergence between adjacent chains, offer a swap-independent perspective. Larger KL divergence implies a greater potential for the colder chain to gain information per swap \cite{Vousden_2015}. However, estimating the KL divergence becomes computationally prohibitive as 
$t \rightarrow \infty$. The Expected Squared Jump Distance (ESJD) addresses this by scaling the expected acceptance rate by the squared temperature difference, $\epsilon^2$. This formulation balances swap efficiency and effectiveness, with larger $\epsilon$ values reflecting more substantial differences between neighboring chains and thus assigning greater significance to successful swaps.

\textbf{Swap Mean Distance}

We propose a metric which is sensitive to both the underlying dynamics of tempered chains and the efficiency of the swap mechanism. For the first, a tempered chain which is sufficiently different from the colder chain should tend to propose states which are far from previously well-explored regions. Thus, an accepted swap to a distant state should be weighted more heavily than to one that is close. Secondly, a rejected proposal should be measured proportionate to its actual impact. Accepting a swap right next to the current state is not much better than rejecting it. This induces a rather natural continuous metric based on distance.

The $m$-mean swap distance $\omega_m$ (referred to as the swap mean-distance) is defined as $$ \omega_m^{(i)}(t) := \frac{1}{m}\sum_{j = 1}^m (d(x^{(i+1)}_t, x^{(i)}_{t-j}))$$ where $d(x,y)$ denotes the Euclidean distance between points $x,y \in \mathbb{R}^n$.

The finite $m > 1$ term represents the "memory" of the chain which persists through swaps. Consider a distribution with three separated modes, for which two chains are initialized at separate modes. If neither have the sufficient energy to hop to the third mode, consistent swapping would still only explore two of the three modes. However, neither the acceptance rate nor the $1$-swap mean-distance metrics would capture this behavior, since immediate swap distances are high and likely. Choosing some $m$ greater than the swap period of $X^{(i)}_t$ and $X^{(i+1)}_t$ ensures that at least some of the states prior to a successful swap are remembered, so that $\omega_m^{(i)}(t)$ is controlled in such scenarios.

We tested several choices of $m$, and found that larger $m$ is typically more strongly correlated with the ACT. This can be explained by the fact that an extended memory captures a larger subset of previously explored states. However, too large an $m$ devalues the influence of the most immediate states. Additionally, it is essential to maintain $m << N$ to control the variance of $\overline{\omega_m^{(i)}}$ estimates. We use $m=50$ in our experiments. 

Empirically, we find strong correlations between the swap mean-distance and ACT. Figure \ref{corr} presents scatterplots for three distinct distributions, namely the multimodal Gaussian, egg-box, and Rosenbrock distributions. The Spearman correlation coefficients for these distributions are found to be are $-0.997, -0.998, -0.856$, respectively, with p-values less than $1e-100$. These findings strongly indicate that swap mean-distance is an effective proxy for ACT. Consequently, the application of swap mean-distance as a reward function is substantiated, with the assumption that ACT generally exhibits non-decreasing behavior in relation to it.

\section{Results}

To evaluate the performance of our algorithm, we conducted experiments on three distributions chosen for their pronounced multimodal or rugged characteristics. Specifically, we employed a mixture of ten 8-dimensional Gaussians, a 5-dimensional egg-box distribution with 243 modes, and the Rosenbrock distribution. Comprehensive details of these distributions are provided in Appendix \ref{dist}.

For each distribution, the algorithm was executed with 15 temperature levels and run for 4000 iterations. We evaluated three reward functions: the swap mean-distance, the expected squared jumping distance, and the negative standard deviation of acceptance rates. Each distribution-reward function pair was subjected to 10 independent trials. After the algorithm's termination, we sampled $N = 10,000$ iterations from the final temperature configuration to compute the average autocorrelation time (ACT) for each trial. These ACT estimates are presented in Table \ref{act_table}.

\begin{table*}[t!]
\label{act_table}
\centering
\caption{ACT estimates for policy gradient algorithm and associated reward functions, and geometric spacing.}
\label{sample-table}
\vskip 0.15in
\begin{small}
\begin{sc}
\begin{tabular}{lcccr}
\toprule
Distribution & Swap Mean-Distance & Acceptance Rate STD & ESJD & Geometric\\
\midrule
Multimodal Gaussian  & \textbf{1.138 $\pm$ 0.06} & 1.222 $\pm$ 0.03 & 2.717 $\pm$ 0.09 & 4.670 $\pm$ 0.05 \\
Egg-box & \textbf{1.097 \ $\pm$ 0.08} & 2.772 $\pm$ 0.22 & 4.817 $\pm$ 0.46 & 9.670 $\pm$ 0.09 \\
Rosenbrock  & \textbf{24.08 \ $\pm$ 4.32} & 43.55 $\pm$ 3.66 & 41.21 $\pm$ 8.54 & 81.32 $\pm$ 8.21 \\
\bottomrule
\end{tabular}
\end{sc}
\end{small}
\vskip -0.1in
\end{table*}

\subsection{ACT Comparison}\label{act_section}

Notably, our policy gradient method consistently outperformed geometric spacing across all reward functions, with the swap mean-distance yielding the lowest ACT for every test distribution. This is consistent with our earlier observation that the swap mean-distance is highly correlated with ACT.

\begin{figure}[h!]
\begin{center}
\centerline{\includegraphics[width=\columnwidth]{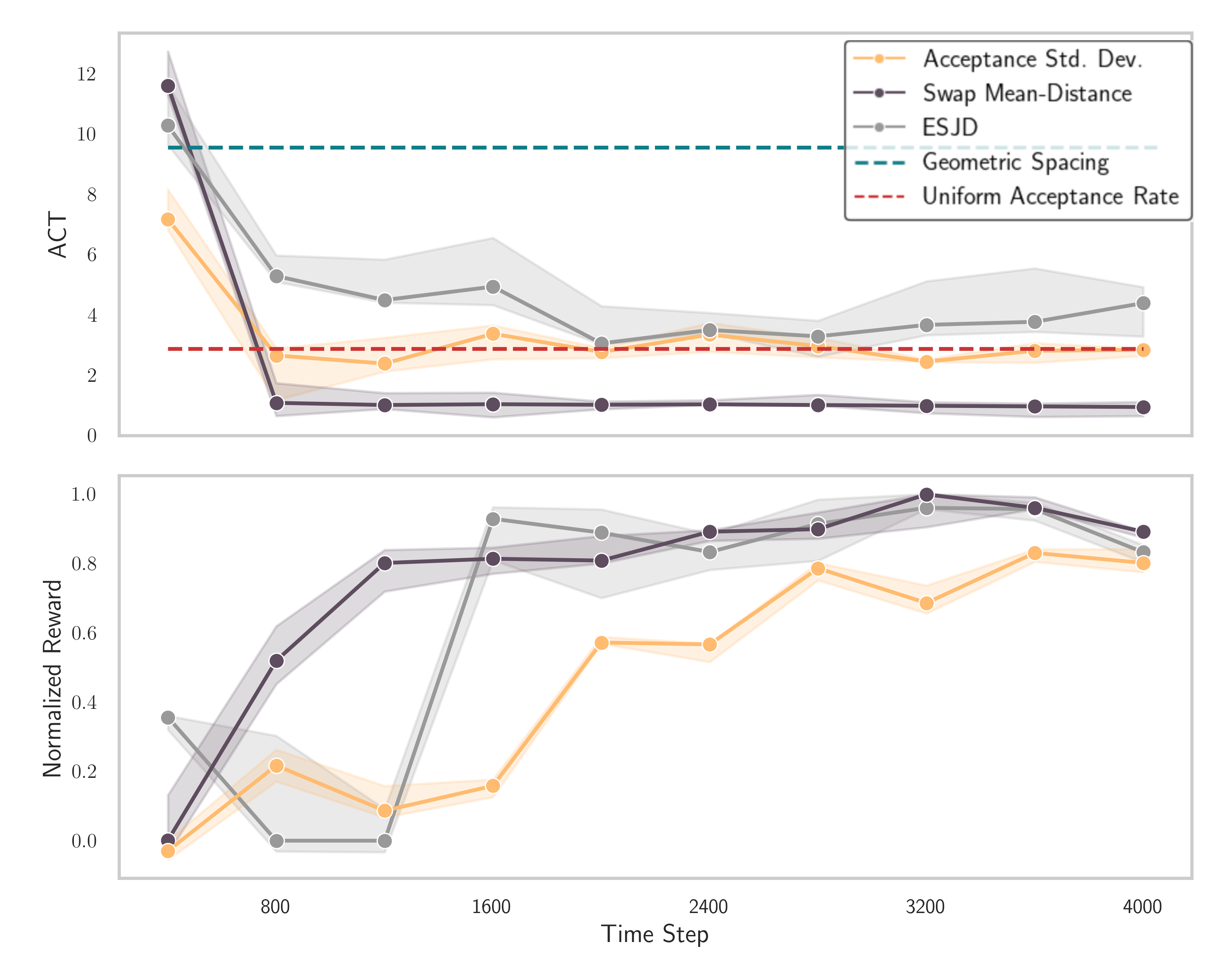}}
\caption{Plot of the ACT over time for geometrically spaced temperatures, uniform acceptance rate \cite{Vousden_2015}, and policy gradient algorithm with different reward functions. Target is the egg-box distribution. Each step represents $400$ iterations of the policy gradient update.}
\label{act}
\end{center}
\vskip -0.5cm
\end{figure}

In Figure \ref{act}, we see that for the egg-box distribution, both the swap mean-distance and acceptance rate standard deviation achieves near-optimal performance after only $800$ iterations. Comparatively, ESJD takes longer to converge and exhibits greater variance across its trials throughout all stages of sampling.

\subsection{Improving on Uniform Acceptance Rates}\label{ars}

The findings from Section \ref{act_section} suggest a need to critically review the longstanding assumption that uniform acceptance rates are inherently optimal. While analogies to physical systems \cite{Sugita1999REMD} offer an appealing heuristic rationale for this assumption, it remains statistically unsubstantiated as a universally optimal strategy for general distributions. The results indicate that more sophisticated approaches to reward shaping could yield improved performance, warranting a deeper exploration of alternative methodologies.

\begin{figure}[h!]
\begin{center}
\hspace*{-0.1cm}
\centerline{\includegraphics[width=\columnwidth]{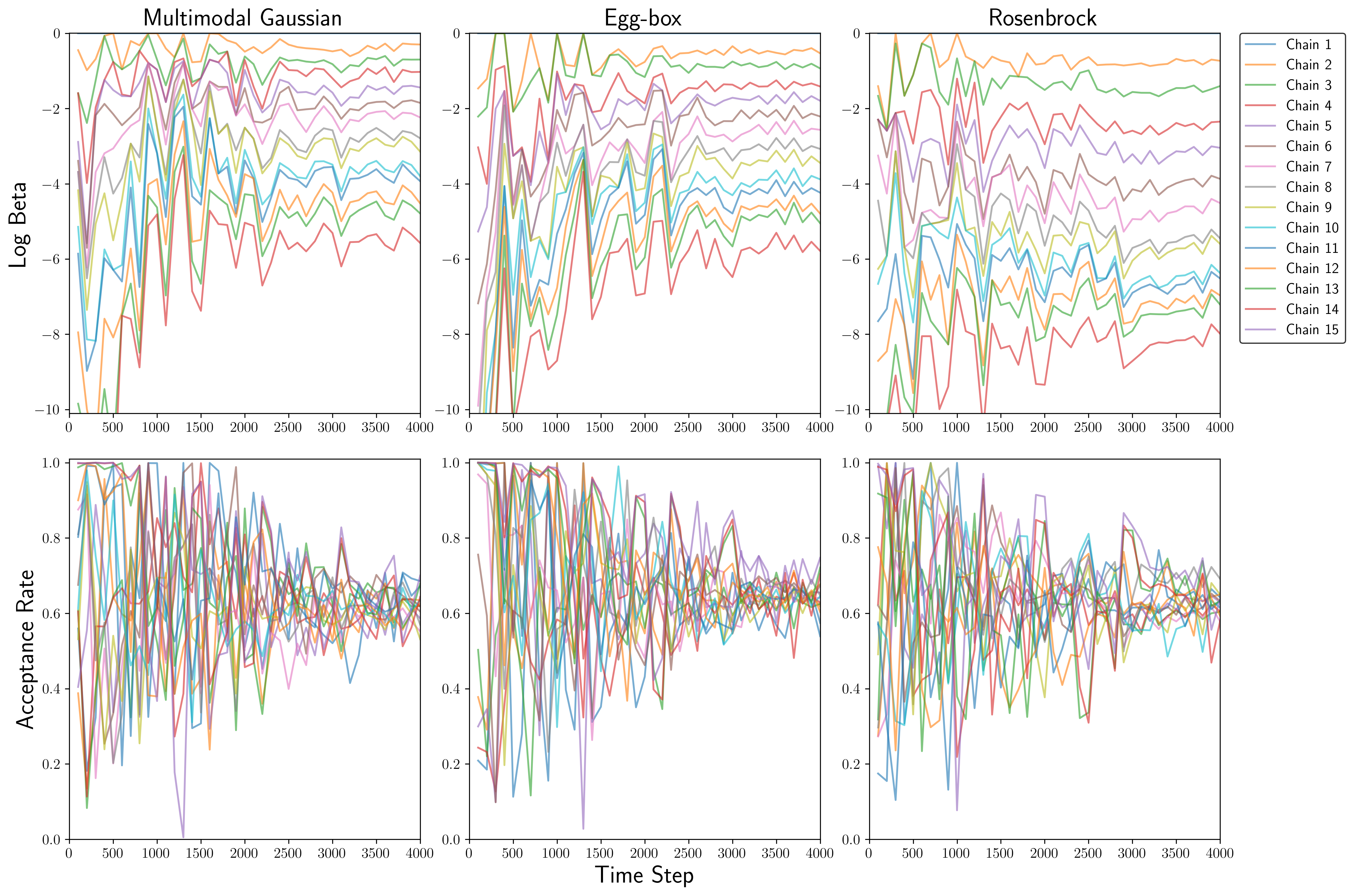}}
\vskip -0.5cm
\caption{Evolution of log $\beta$ and acceptance rates over $4000$ update steps. Target is the egg-box distribution. Data is thinned by a factor of $100$.}
\label{ar}
\end{center}
\vskip -0.1cm
\end{figure}

As a baseline, we demonstrate that uniform acceptance rates is achievable via policy gradients by setting our reward function to the scaled negative standard deviation of the acceptance rates. As illustrated in Fig. \ref{ar}, the acceptance rates of all chains converge uniformly. 

We emphasize that improvement over the uniform acceptance rate paradigm was achieved simply by modifying the reward function of the policy gradient algorithm. This method is applicable to nearly any desired temperature configuration for which a respective reward can be shaped, opening doors for understanding of the efficiency of its swapping mechanism.

However, note that because the swap-mean distance metric is based only on the coldest chain, it is not particularly sensitive to the dynamics of hot chains. For example, Figure \ref{box} demonstrates that across all the trials for the egg-box distribution the first four temperatures When $M$ is large, it is possible that hotter chains are slow to converge to optimal temperatures. It may be advantageous to use a weighted combination of the swap mean-distance metric and a function of acceptance rates (e.g. minimum, harmonic mean, standard deviation, etc.) as the reward function, or imposing penalties on bottleneck pairs of adjacent chains whose acceptance rate is close to $0$. 

\section*{Conclusion}

In this paper, we present a policy gradient-based optimization method for tuning the temperature ladder in parallel tempering MCMC. Our algorithm is supported by theoretical convergence results and empirical experiments demonstrating its effectiveness. We introduce a novel optimization metric that reflects the dynamics of the swap mechanism in parallel tempering. Our findings reveal a strong relationship between this metric and the sampler's ACT. By integrating this metric into our policy gradient framework, we achieve substantially lower ACTs compared to those obtained using conventional uniform acceptance rate strategies.

\section*{Acknowledgements}
We thank Prof. Eric Laber and Dr. Zachary Moore for their generous input into this project and help with ideas from Reinforcement Learning. Some of this work was inspired by similar work with the above two researchers on other MCMC algorithms.






\bibliography{main}
\bibliographystyle{icml2024}

\newpage
\appendix
\onecolumn
\section{Adaptive Affine Algorithm Details.}

In Sections \ref{rw} and \ref{act_section}, we reference an adaptive parallel tempering algorithm \cite{Vousden_2015} that results in uniform acceptance rates, and the ensemble-based affine invariant sampler \cite{goodman2010} on which it is built. We provide details on these algorithms.

\subsection{Affine Invariant Ensemble Sampler.}
The core idea behind the affine invariant sampler is that it can use context information from other chains in an ensemble of chains to perform well under affine transformations of the parameter space. An affine transformation involves linear transformations such as scaling, translation, rotation, and shearing. Metropolis-Hastings proposal distributions have trouble navigating these features, but the ensemble sampler makes use of the "stretch move" to overcome this issue.

\begin{algorithm}
\caption{Affine Invariant Ensemble Sampler}\label{alg:aies}
\begin{algorithmic}[1]
\INPUT Initial positions of walkers $\{\mathbf{x}_i^{(0)}\}_{i=1}^{N}$, target distribution $\pi(\mathbf{x})$, stretch move parameter $a$

\STATE Initialize the ensemble of walkers $\{\mathbf{x}_i^{(0)}\}_{i=1}^{N}$
\FOR{$t = 0$ to $T-1$}
    \FOR{each walker $i$}
        \STATE Randomly select another walker $j \neq i$ from the ensemble
        \STATE Draw $z$ from the distribution $g(z) \propto \frac{1}{\sqrt{z}}$ defined over $\left[\frac{1}{a}, a\right]$
        \STATE Propose a new position $\mathbf{x}_i' = \mathbf{x}_j + z(\mathbf{x}_i^{(t)} - \mathbf{x}_j)$
        \STATE Compute acceptance probability $\alpha = \min\left(1, z^{d-1} \frac{\pi(\mathbf{x}_i')}{\pi(\mathbf{x}_i^{(t)})}\right)$
        \STATE Draw a uniform random number $u \sim \mathcal{U}(0, 1)$
        \IF{$u < \alpha$}
            \STATE Accept the proposal: $\mathbf{x}_i^{(t+1)} = \mathbf{x}_i'$
        \ELSE
            \STATE Reject the proposal: $\mathbf{x}_i^{(t+1)} = \mathbf{x}_i^{(t)}$
        \ENDIF
    \ENDFOR
\ENDFOR
\OUTPUT $\{\mathbf{x}_i^{(t)}\}_{i=1}^{N}$
\end{algorithmic}
\end{algorithm}

In context of parallel tempering, this method is adapted so that an ensemble sampler is constructed at each temperature level. Each ensemble only considers the samplers in its own temperature level when performing the stretch move. Independently, after each stretch move, a random permutation of swaps are proposed between every sampler of ensembles in adjacent temperature levels. In our experiments, each ensemble is composed of $16$ samplers.

\subsection{Adaptive Parallel Tempering Ensemble.}
The adaptive parallel tempering algorithm constructs an ensemble sampler for each temperature in the ladder and updates the log differences $S_i := \log (T_{i+1} - T_i)$ between temperatures based on neighboring acceptance rates: $$S_i(t+1) = S_i(t) + \kappa(t)[A_i(t) - A_{i+1}(t)].$$ The function $\kappa(t)$ is a hyperbolic decay function which dampens the amplitude of adaptations over time to ensure convergence. This updating mechanism is quite intuitive—each $S_i$ is adjusted to close the gap between two chains with relatively low acceptance rates, or repel if they have high acceptance rates.

\section{Toy Distributions.} \label{dist}
We provide specifics on the toy distributions we used to test our algorithm. The following three distributions all exhibit multimodality to varying degrees. The Rosenbrock function additionally possesses a narrow, curved valley between modes, testing the algorithm's capability to navigate extreme topographies.

\textbf{Multimodal Gaussian Distribution.}
Given parameters $\{w_i, \mu_i, \Sigma_i\}_{1 \leq i \leq n}$, the Multimodal Gaussian is a mixture of Gaussian distributions, which has likelihood

$$L(\theta) \propto \sum_{i=1}^n w_i \cdot p(\theta | \mu_i, \Sigma_i)$$ where $$p(\mathbf{x}|\mathbf{\mu}, \mathbf{\Sigma}) = \frac{1}{(2\pi)^{k/2} |\mathbf{\Sigma}|^{1/2}} \exp \left( -\frac{1}{2} (\mathbf{x} - \mathbf{\mu})^\top \mathbf{\Sigma}^{-1} (\mathbf{x} - \mathbf{\mu}) \right).$$ In our experiments, $n=10$ and each $\mu_i$ is chosen uniformly at random in the interval $[-1, 1]$. Furthermore, $\Sigma_i = \sigma_i I$ where $\sigma_i \sim [0.01, 0.3]$ uniformly. The parameter space is restricted to $[-2, 2]^8$.

\textbf{Egg-box Distribution.}
The egg-box distribution is a popular test distribution in machine learning and optimization because it presents several isolated modes. Its density is
$$L(\theta) \propto \left(\frac{1}{2}\prod_{i=1}^d \cos{\theta_i} + \frac{1}{2}\right)^\beta.$$

Under large $\beta$, these modes look locally Gaussian and do not overlap unless the distribution is sufficiently tempered. We take $\beta = 1000$ in our experiments and restrict the parameter space to $[-3\pi/2, 3\pi/2]^5$.

\textbf{Rosenbrock Distribution.}
The Rosenbrock distribution contains two modes with extremely narrow, curved valleys which are difficult to traverse. Its density is given by
$$L(x, y) \propto \left( \frac{1}{c + f(x, y)} + \frac{1}{c + f(-x, y)} \right)^\beta$$ where $$f(x,y) = (a-x^2)^2 + b(y-x^2)^2.$$ In our experiments, we take $a=4, b=1, c=0.1, \beta=1000$.

\newpage

\section{Experiments Extended.}

\subsection{Markov Chain Convergence in Log-Likelihood.}
\begin{figure}[ht]
\vskip -0.2in
\begin{center}
\centerline{\includegraphics[width=0.9\columnwidth]{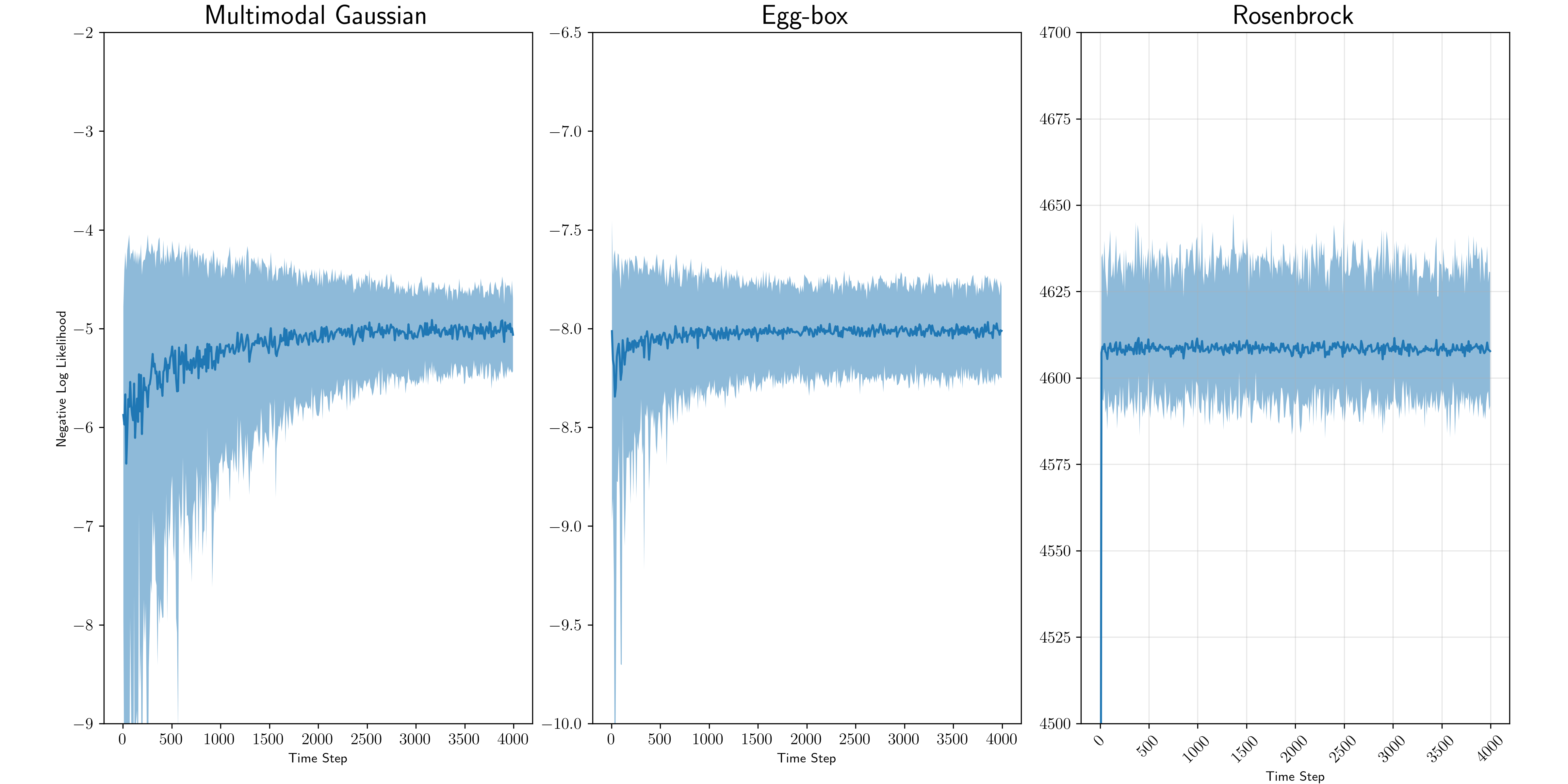}}
\caption{Averaged negative log likelihoods of the samples each time step of the algorithm. The shaded region indicates a $95$\% confidence interval over $10$ trials.}
\label{a:corr}
\end{center}
\vskip -0.2in
\end{figure}

When sampling from complex distributions, a key challenge is to ensure that the Markov chain converges to the target distribution within a finite number of iterations. To evaluate the convergence of our algorithm, we plot the trajectory of the negative unnormalized log-likelihood over the course of $4,000$ iterations in Figure \ref{a:corr}. The negative log-likelhood functions appear to approach stationarity in all three distributions, suggesting convergence of the Markov chains.

\subsection{Swap Mean Distance Correlation with ACT.}
\begin{figure}[ht]
\vskip 0.2in
\begin{center}
\centerline{\includegraphics[width=\columnwidth]{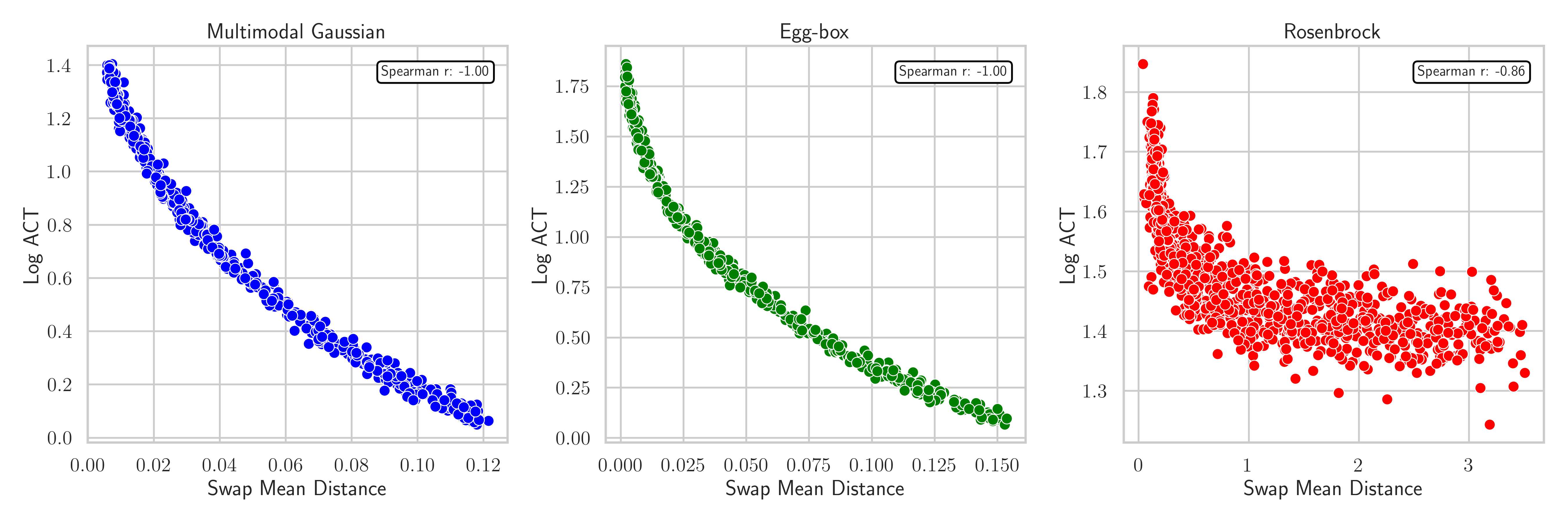}}
\caption{Scatter plot of swap mean-distance against ACT.}
\label{corr}
\end{center}
\vskip -0.2in
\end{figure}

We plot the scatterplot between swap mean-distance and estimated ACT for three distributions in Figure \ref{corr}. Data is generated by randomly selecting $1000$ temperature ladders for each $m$ value and estimating the average swap mean-distance and ACT with $N=1000$ MCMC iterations.

\subsection{Distribution of Final Betas.}
\begin{figure}[ht]
\vskip 0.2in
\begin{center}
\centerline{\includegraphics[width=\columnwidth]{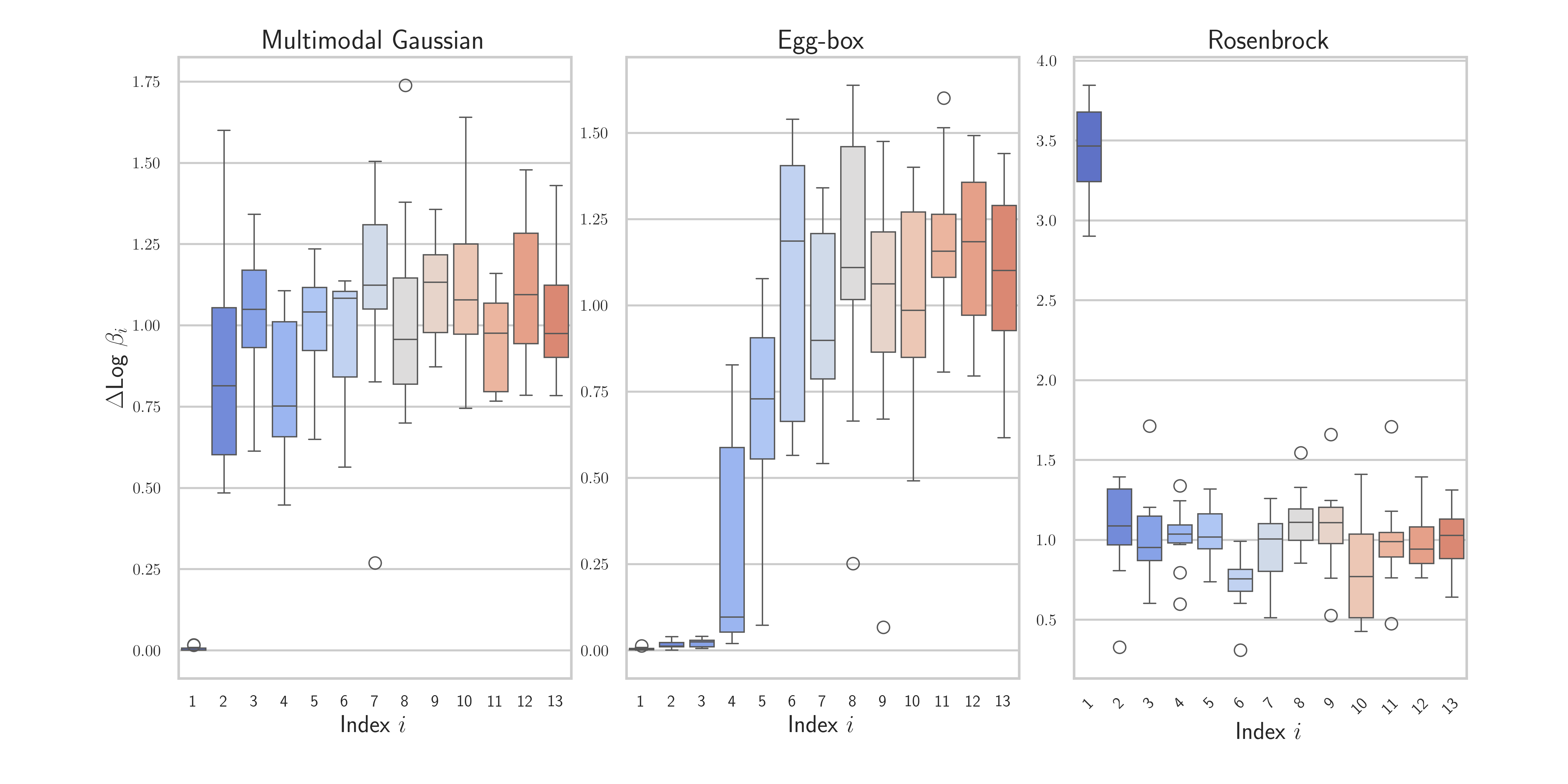}}
\caption{Box plot of final temperatures after $4000$ iterations with the swap mean-distance reward function.}
\label{box}
\end{center}
\vskip -0.2in
\end{figure}

Though our adaptive algorithm is ergodic, the temperature ladder to which it converges is not deterministic. This variability is particularly pronounced when optimizing non-convex reward functions, such as the swap mean-distance. Notably, the choice of random seed can result in substantial divergence in the final temperature configuration. Figure \ref{box} illustrates this phenomenon, presenting a box plot that captures the distribution of final parameters in terms of $\Delta \log \beta_i$ when the swap mean-distance reward function is employed.

\end{document}